\documentclass[11pt, a4paper]{article}
\usepackage[utf8]{inputenc}
\usepackage[T1]{fontenc}
\usepackage{geometry}
\usepackage{booktabs}
\usepackage{amsmath, amssymb}
\usepackage{graphicx}
\usepackage{hyperref}
\usepackage{longtable}
\usepackage{caption}
\usepackage{natbib}
\usepackage{xcolor}
\usepackage{setspace}

\usepackage{colortbl}

\usepackage{pgfplots}
\pgfplotsset{compat=1.18}
\usepgfplotslibrary{fillbetween}
\usepackage{tikz}

\title{Status of the $S_8$ Tension: A 2026 Review of Probe Discrepancies}
\author{
     Ioannis Pantos and Leandros Perivolaropoulos \\ 
    \small Department of Physics, University of Ioannina \\
    \small Ioannina 45110, Greece
}
\date{\today}

\begin{document}

\maketitle

\begin{abstract}

The parameter $S_8 \equiv \sigma_8 (\Omega_m/0.3)^{0.5}$ quantifies the amplitude of matter density fluctuations. A persistent discrepancy exists between early-universe CMB observations and late-universe probes. This review assesses the ``$S_8$ tension'' against a new 2026 baseline: a unified ``Combined CMB'' framework incorporating Planck, ACT DR6, and SPT-3G. This combined analysis yields $S_8 = 0.836^{+0.012}_{-0.013}$, providing a higher central value and reduced uncertainties compared to Planck alone. Compiling measurements from 2019--2026, we reveal a striking bifurcation: DES Year 6 results exhibit a statistically significant tension of $2.4\sigma$--$2.7\sigma$ in $S_8$ \citep{DESY6}, whereas KiDS Legacy results demonstrate statistical consistency at $<1\sigma$ \citep{Wright2025}. We examine systematic origins of this dichotomy, including photometric redshift calibration, intrinsic alignment modeling, and shear measurement pipelines. We further contextualize these findings with cluster counts (where eROSITA favors high values while SPT favors low), galaxy-galaxy lensing, and redshift-space distortions. The heterogeneous landscape suggests survey-specific systematic effects contribute substantially to observed discrepancies, though new physics beyond $\Lambda$CDM cannot be excluded.
\end{abstract}

\tableofcontents
\newpage

%==============================================================================
\section{Introduction}
%==============================================================================

\subsection{The Emergence of Precision Cosmology}

The past three decades have transformed cosmology from a largely qualitative discipline into a precision science. The establishment of the $\Lambda$CDM concordance model---characterized by a cosmological constant $\Lambda$, cold dark matter, baryonic matter, and a nearly scale-invariant spectrum of primordial perturbations---represents one of the most significant achievements in modern physics. This model successfully describes observations spanning fourteen orders of magnitude in scale, from the acoustic peaks in the CMB power spectrum to the distribution of galaxies in the cosmic web.

The parameters of the $\Lambda$CDM model have been constrained with remarkable precision through a diverse array of cosmological probes. Primary among these are observations of the CMB, which encode information about the universe at the epoch of recombination approximately 380,000 years after the Big Bang. The \textit{Planck} satellite, operational from 2009 to 2013, provided the definitive space-based measurement of CMB temperature and polarization anisotropies \citep{Planck2020}, achieving cosmic variance-limited measurements over a substantial fraction of the sky and across a wide range of angular scales.

The concordance between independent cosmological probes has been a cornerstone of modern cosmology's success. However, as measurement precision has improved, subtle but persistent discrepancies have emerged. The most prominent of these---the Hubble tension between early- and late-universe determinations of $H_0$---has garnered substantial attention. Less widely discussed but equally intriguing is the tension in measurements of the amplitude of matter clustering, parameterized by $S_8$.

\subsection{The \texorpdfstring{$S_8$}{S8} Parameter: Definition and Significance}

Among the parameters characterizing the $\Lambda$CDM model, the amplitude of matter density fluctuations holds particular importance for understanding structure formation. This amplitude is conventionally parameterized by $\sigma_8$, defined as the root-mean-square of the linear matter density field when smoothed with a top-hat filter of radius $8 \, h^{-1}$ Mpc:
\begin{equation}
    \sigma_8^2 = \int_0^\infty \frac{k^2 P(k)}{2\pi^2} \left| W(kR) \right|^2 \, dk,
\end{equation}
where $P(k)$ is the linear matter power spectrum, $W(kR)$ is the Fourier transform of the top-hat window function, and $R = 8 \, h^{-1}$ Mpc. The choice of $8 \, h^{-1}$ Mpc is historical, corresponding roughly to the scale at which the variance of galaxy counts in spheres equals unity.

Late-universe probes of large-scale structure, particularly weak gravitational lensing surveys, exhibit a characteristic degeneracy between $\sigma_8$ and the matter density parameter $\Omega_m$. This degeneracy arises because cosmic shear measurements are sensitive to the integrated matter power spectrum along the line of sight, weighted by geometric lensing kernels. The resulting constraints trace out elongated contours in the $\sigma_8$--$\Omega_m$ plane, approximately following curves of constant $S_8 \equiv \sigma_8 (\Omega_m/0.3)^{\alpha}$, where the exponent $\alpha$ depends on the specific observable and redshift distribution but typically falls in the range $0.4$--$0.6$. The conventional choice $\alpha = 0.5$ has been adopted as a standard, yielding:
\begin{equation}
    S_8 \equiv \sigma_8 \sqrt{\frac{\Omega_m}{0.3}}.
\end{equation}

This combination is particularly well-constrained by weak lensing measurements and serves as the primary diagnostic for comparing early- and late-universe determinations of the clustering amplitude. As emphasized by \citet{AmonEfstathiou2022}, the interpretation of $S_8$ constraints requires careful attention to the scales being probed and potential systematic effects that may differ between linear and nonlinear regimes.

\subsection{Historical Context: The Evolution of the \texorpdfstring{$S_8$}{S8} Tension}

The $S_8$ tension first emerged as a statistically significant discrepancy following the \textit{Planck} 2015 data release, which reported $S_8 = 0.851 \pm 0.024$ under the assumption of a flat $\Lambda$CDM cosmology. Contemporary weak lensing analyses from the Canada-France-Hawaii Telescope Lensing Survey (CFHTLenS) and early results from KiDS consistently favored lower values, typically in the range $S_8 \approx 0.75$--$0.78$. The statistical significance of this discrepancy ranged from $2\sigma$--$3\sigma$  depending on the specific datasets and analysis choices employed.

The \textit{Planck} 2018 final data release refined the CMB-based constraint to $S_8 = 0.832 \pm 0.013$ \citep{Planck2020}, a modest downward revision that nonetheless maintained tension with the majority of weak lensing measurements. Throughout the period 2018--2024, multiple independent weak lensing analyses---including KiDS-1000 \citep{Asgari2021}, DES Year 3 \citep{Abbott2022} $\sim 2.6\sigma$ relative to Planck, and HSC Year 1 \citep{Hikage2019} $\sim 1.6\sigma$---reported $S_8$ values systematically below the 
\textit{Planck} prediction, with discrepancies ranging 
from ${\sim}1.6\sigma$ to ${\sim}3\sigma$, reaching 
${\sim}3.5\sigma$ for RSD compilations 
\citep{Benisty2021}.

The persistence of this tension across multiple independent surveys lent credibility to the hypothesis that it might reflect genuine new physics rather than systematic errors. Proposed explanations have included modifications to gravity \citep{Boiza2025,Souza2025,Terasawa2025,HuSawicki2007}, decaying dark matter \citep{Tanimura2023}, interacting dark energy \citep{Sabogal2024}, and various extensions to the neutrino sector. However, as we shall discuss, the heterogeneous nature of the current tension landscape complicates such interpretations.

\subsection{The 2026 Paradigm Shift: Combined CMB as the New Baseline}
\label{sec:combined_cmb}

The cosmological landscape underwent a significant transformation in early 2026 with the maturation of ground-based CMB experiments to a precision level rivaling and complementing \textit{Planck}. The Atacama Cosmology Telescope (ACT) Data Release 6 \citep{Louis2025} and the South Pole Telescope 3rd Generation (SPT-3G) \citep{Camphuis2025} now provide high-resolution measurements of the CMB damping tail at multipoles $\ell > 2000$, where \textit{Planck}'s angular resolution becomes limiting. These ground-based experiments also offer independent measurements of CMB lensing \citep{Qu2024ACTDR6}, providing crucial cross-checks on systematic uncertainties.

The combination of \textit{Planck} 2018, ACT DR6, and SPT-3G data---referred to throughout this
review as the ``Combined CMB'' dataset---yields a constraint of $S_8 = 0.836^{+0.012}_{-0.013}$
\citep{DESY6}. This represents several important developments relative to the
\textit{Planck}-only baseline:
\begin{enumerate}
    \item The central value has shifted upward by approximately $0.004$, or
    roughly $0.3\sigma$ in the \textit{Planck} error budget.
    \item The uncertainties remain comparable to the \textit{Planck}-only
    case, with a slight reduction in the upper uncertainty from $+0.013$
    to $+0.012$.
    \item The combination of independent experiments with different
    systematic error profiles enhances confidence in the robustness of
    the result.
\end{enumerate}

The Combined CMB baseline $S_8 = 0.836^{+0.012}_{-0.013}$ was derived by the DES collaboration \citep{DESY6} from a joint analysis of the \textit{Planck} 2018, ACT-DR6, and SPT-3G primary CMB likelihoods, excluding CMB lensing. To our knowledge, no independent CMB-only joint analysis of this specific combination has been published separately.

This review adopts the Combined CMB baseline as the primary reference against which late-universe measurements are compared. Where relevant, we also discuss tensions relative to the \textit{Planck} 2018 baseline for historical context and comparison with earlier literature.

%==============================================================================
\section{Methodology: Quantifying Cosmological Tensions}
%==============================================================================

\subsection{Statistical Framework for Tension Assessment}

Before presenting the compilation of $S_8$ measurements, it is essential to establish the statistical framework employed for quantifying tensions between datasets. Several approaches exist in the literature, each with distinct advantages and limitations.

The simplest and most commonly employed metric is the difference in mean values normalized by the combined uncertainty:
\begin{equation}
    n_\sigma = \frac{|S_8^{\rm (A)} - S_8^{\rm (B)}|}{\sqrt{\sigma_{\rm A}^2 + \sigma_{\rm B}^2}},
\end{equation}
where $S_8^{\rm (A)}$ and $S_8^{\rm (B)}$ are the central values from datasets A and B, and $\sigma_{\rm A}$ and $\sigma_{\rm B}$ are the corresponding uncertainties. This metric assumes Gaussian posteriors and statistical independence between the datasets.

More sophisticated tension metrics have been developed to overcome limitations of 
the simple $n_\sigma$ approach. \citet{Handley2019} introduced the Suspiciousness 
statistic, a prior-insensitive Bayesian method that accounts for the full 
multi-dimensional parameter space. \citet{Troster2020} employed both the Bayesian 
evidence ratio and Suspiciousness statistic to quantify the $S_8$ discrepancy, 
finding a $2.1 \pm 0.3\sigma$ tension between combined BOSS galaxy clustering 
and KV450 weak lensing data and Planck CMB constraints---broadly consistent with 
simpler metrics when posteriors are approximately Gaussian.

In cases where posteriors are significantly non-Gaussian or exhibit complex degeneracies, these methods may yield tension assessments that differ from the simple $n_\sigma$ metric. Throughout this review, we report tensions using the $n_\sigma$ metric for consistency and ease of comparison, while noting cases where more detailed statistical analyses have been performed and yield different conclusions.

\subsection{Treatment of Asymmetric Uncertainties}

Many $S_8$ measurements report asymmetric uncertainties, reflecting 
non-Gaussian posterior distributions. For such cases, we adopt the 
convention of using the uncertainty in the direction of the 
discrepancy when computing $n_\sigma$. Specifically, when comparing 
a measurement $S_8 = x^{+\sigma_1}_{-\sigma_2}$ against the 
Combined CMB baseline ($S_8 = 0.836^{+0.012}_{-0.013}$), we use 
$\sigma_2$ and the baseline lower uncertainty if $x < 0.836$, 
and $\sigma_1$ and the baseline upper uncertainty if $x > 0.836$.

\subsection{Correlation Between Measurements}

A significant complication in assessing the global significance of the $S_8$ tension arises from correlations between different measurements. Surveys covering overlapping sky areas will exhibit correlated cosmic variance contributions. Different analyses of the same raw data (e.g., varying analysis choices within a single survey collaboration) are even more strongly correlated. Throughout this review, we identify cases where measurements are derived from the same underlying data and should not be treated as independent constraints.

The joint analysis of KiDS and DES data by \citet{Kilo-DegreeSurvey:2023gfr} represents an important step toward properly accounting for such correlations when combining constraints from different surveys.

%==============================================================================
\section{The Combined CMB Baseline: Technical Details}
%==============================================================================

\subsection{Component Datasets}

The Combined CMB baseline integrates three primary datasets, each contributing distinct information about the primordial universe.

\subsubsection{Planck 2018}

The \textit{Planck} satellite observed the CMB in nine frequency bands spanning 30--857 GHz, enabling robust separation of the CMB signal from Galactic and extragalactic foregrounds. The 2018 legacy data release \citep{Planck2020} includes temperature (TT) and polarization (TE, EE) power spectra spanning multipoles $2 \leq \ell \leq 2500$ for temperature and $2 \leq \ell \leq 2000$ for polarization. The low-$\ell$ likelihood ($\ell < 30$) constrains the optical depth to reionization $\tau$, which is partially degenerate with $A_s$ (the amplitude of primordial scalar perturbations) and hence affects $\sigma_8$ inference. The high-$\ell$ likelihood constrains the shape and amplitude of the acoustic peaks, providing sensitivity to $\Omega_m$, $\Omega_b h^2$, $H_0$, and $n_s$.

Under the assumption of flat $\Lambda$CDM with minimal neutrino mass ($\sum m_\nu = 0.06$ eV), \textit{Planck} 2018 alone yields $S_8 = 0.832 \pm 0.013$ \citep{Planck2020}.

\subsubsection{ACT Data Release 6}

The Atacama Cosmology Telescope is a 6-meter telescope located in the Atacama Desert of Chile at an elevation of 5190 meters. Its high angular resolution (approximately 1.4 arcminutes at 150 GHz) enables precise measurements of the CMB damping tail at $\ell > 2000$, where \textit{Planck}'s beam dilutes the signal. ACT DR6 \citep{Louis2025} includes temperature and polarization power
spectra from observations conducted between 2017 and 2022, with maps
covering approximately 45\% of the sky ($\sim$19\,000\,deg$^2$) of which $\sim$25\% is used
for the power spectrum analysis after masking the Galactic plane and
bright extragalactic sources.

The damping tail measurements are particularly valuable for constraining the effective number of relativistic species $N_{\rm eff}$ and the primordial helium abundance $Y_p$, parameters that affect the shape of the power spectrum at small scales. ACT DR6 also provides an independent measurement of CMB lensing \citep{Qu2024ACTDR6}, which directly probes the integrated matter distribution along the line of sight and yields constraints on $S_8$ that are complementary to the primary CMB anisotropies.

\subsubsection{SPT-3G}

The South Pole Telescope is a 10-meter telescope located at the Amundsen-Scott South Pole Station. Its location provides exceptional atmospheric stability, enabling deep integrations over a contiguous 1500 deg$^2$ survey field. SPT-3G, the third-generation camera, employs approximately 16,000 polarization-sensitive bolometers operating at 95, 150, and 220 GHz.

SPT-3G \citep{Camphuis2025} contributes high signal-to-noise measurements of the E-mode polarization power spectrum and provides the most precise ground-based measurements of the damping tail at $\ell > 3000$. The combination of SPT-3G with \textit{Planck} and ACT yields improved constraints on parameters affecting the high-$\ell$ spectrum, including the damping scale and the amplitude of gravitational lensing of the CMB.

\subsection{Combination Methodology}

The combination of \textit{Planck}, ACT, and SPT data requires careful treatment of several effects. Different experiments observe overlapping regions of sky, introducing covariance between their measurements due to shared cosmic variance. This covariance must be properly modeled to avoid underestimating parameter uncertainties. Additionally, each experiment employs distinct foreground models, beam characterization procedures, and calibration strategies. Joint analyses must either marginalize over inter-experiment calibration differences or demonstrate consistency when adopting unified calibration.

The Combined CMB analysis presented in association with DES Y6 \citep{DESY6} employs a likelihood that accounts for cross-correlations between experiments while marginalizing over residual calibration uncertainties. The resulting constraint $S_8 = 0.836^{+0.012}_{-0.013}$ is robust to reasonable variations in analysis choices.

An independent but related constraint comes from CMB lensing. The combined ACT, SPT, and \textit{Planck} lensing analysis \citep{Qu2026} constructs a joint Gaussian likelihood from the CMB lensing bandpowers of ACT DR6, \textit{Planck} PR4, and SPT-3G. When supplemented with DESI DR2 BAO data to break the degeneracy between $\sigma_8$ and $\Omega_m$, they obtain a competitive constraint on the parameter combination best measured by cosmic shear, $S_8 \equiv \sigma_8(\Omega_m/0.3)^{0.5} = 0.828 \pm 0.012$. This CMB lensing constraint probes the matter distribution at intermediate redshifts ($z \approx 0.9$--$5$) and provides an important bridge between the recombination epoch and the late universe probed by galaxy surveys.

%==============================================================================
\section{Late-Universe Probes: Physical Principles and Systematic Uncertainties}
%==============================================================================

\subsection{Weak Gravitational Lensing}

Weak gravitational lensing refers to the coherent distortion of background galaxy images by intervening large-scale structure. The observable quantity is the cosmic shear field $\gamma$, which can be decomposed into E-mode and B-mode components analogous to CMB polarization. In the absence of systematic effects, cosmological lensing produces only E-modes; non-zero B-modes indicate residual systematics from PSF modeling, detector effects, or other sources.

The cosmic shear two-point correlation function (or equivalently, the shear power spectrum) is related to the matter power spectrum through a weighted projection:
\begin{equation}
    C_\ell^{ij} = \int_0^{\chi_H} d\chi \, \frac{W^i(\chi) W^j(\chi)}{\chi^2} P\left(\frac{\ell + 1/2}{\chi}, z(\chi)\right),
\end{equation}
where $W^i(\chi)$ is the lensing efficiency kernel for source redshift bin $i$, $\chi$ is the comoving distance, and $P(k, z)$ is the nonlinear matter power spectrum. The Limber approximation has been employed here, which is accurate for $\ell \gtrsim 10$.

The dependence on $S_8$ enters primarily through the amplitude of the matter power spectrum, while the shape of the shear correlation function constrains the combination of $\Omega_m$ and the spectral index $n_s$. The characteristic degeneracy direction in the $\sigma_8$--$\Omega_m$ plane arises from the geometric weighting by the lensing kernels, which are sensitive to both the amplitude and redshift distribution of source galaxies.

\subsubsection{Systematic Uncertainties in Cosmic Shear}

Several systematic effects can bias cosmic shear measurements and hence $S_8$ inference. These have been extensively discussed in the context of DES \citep{Yamamoto:2025jsd}, KiDS \citep{Wright2025,Li2023KiDSPhotoz}, and HSC \citep{Dalal2023,Hikage2019}.

\textit{Shear measurement bias}: The shear field must be estimated from noisy galaxy images with complex morphologies. Multiplicative biases $m$ (defined such that the estimated shear $\hat{\gamma} = (1+m)\gamma^{\rm true}$) directly propagate to $S_8$ as $\Delta S_8 / S_8 \approx m$. Current surveys aim for $|m| < 0.01$, but achieving this requires sophisticated calibration using image simulations that accurately reproduce the properties of observed galaxies. The DES Y6 analysis employs the METADETECTION algorithm \citep{Yamamoto:2025jsd}, which achieves multiplicative bias uncertainties below 0.5\%.

\textit{Photometric redshift errors:} Cosmic shear analyses rely on 
photometric redshifts to assign galaxies to tomographic bins. Biases in 
the mean redshift of each bin shift the inferred $S_8$, with typical 
sensitivities of $\Delta S_8 / \Delta\langle z\rangle \approx -0.4\, S_8$ 
per bin. The KiDS-Legacy analysis \citep{Wright2025} employs a 
sophisticated self-organizing map (SOM) methodology combined with 
spectroscopic calibration samples to achieve redshift distribution 
uncertainties at the sub-percent level in mean redshift.

\textit{Intrinsic alignments}: Galaxies are not randomly oriented tracers of the shear field; they are physically connected to the large-scale structure in which they form. Tidal interactions with the surrounding density field induce coherent alignments that mimic the cosmic shear signal. These contributions manifest as two distinct terms: ``intrinsic-intrinsic'' (II) correlations, arising between physically close galaxies aligned with the same local tidal field, and     ``gravitational-intrinsic'' (GI) correlations, which occur when a foreground structure aligns a foreground galaxy while simultaneously lensing a background galaxy. The physical origin of these alignments depends on galaxy type: pressure-supported elliptical (red) galaxies are primarily aligned through tidal stretching, a mechanism well-described by the Nonlinear Alignment (NLA) model. In contrast, angular-momentum-supported spiral (blue) galaxies are influenced by tidal torquing, which requires higher-order descriptions such as the Tidal Alignment and Tidal Torquing (TATT) model utilized in recent analyses like DES Y6. Unmodeled IA typically suppresses the observed shear signal (dominated by the negative GI term), potentially biasing inferred $S_8$ values low if not properly accounted for. However, the choice of IA model involves a critical trade-off: while simple models like NLA may be insufficient for deep surveys mixing galaxy populations, highly flexible models like TATT introduce broad priors that can degrade constraining power and potentially absorb non-IA physical signals, such as baryonic suppression.

\textit{Baryon feedback}: The nonlinear matter power spectrum is modified on small scales ($k > 1 \, h$ Mpc$^{-1}$) by baryonic processes including gas cooling, star formation, and AGN feedback. The FLAMINGO hydrodynamical simulations \citep{Schaye2023,McCarthy2023} have demonstrated that these effects can suppress power by up to 20--30\% at $k \sim 10 \, h$ Mpc$^{-1}$, potentially biasing $S_8$ constraints if not properly accounted for. Current cosmic shear analyses marginalize over feedback uncertainties using parameterized models calibrated against simulations.

\subsection{Galaxy Clustering and the \texorpdfstring{3$\times$2pt}{3x2pt} Framework}

Galaxy clustering measurements probe the matter distribution through the positions of galaxies, which trace the underlying density field with a bias $b$ that depends on galaxy type, luminosity, and redshift. The galaxy angular power spectrum $C_\ell^{gg}$ is related to the matter power spectrum by:
\begin{equation}
    C_\ell^{gg,ij} = \int d\chi \, \frac{n^i(\chi) n^j(\chi)}{\chi^2} b^i(z) b^j(z) P\left(\frac{\ell + 1/2}{\chi}, z\right),
\end{equation}
where $n^i(\chi)$ is the comoving number density distribution of lens galaxies in bin $i$.

The combination of cosmic shear, galaxy clustering, and galaxy-galaxy 
lensing (the cross-correlation between lens positions and source shears) 
constitutes the ``$3\times2$pt'' analysis framework employed by DES 
\citep{DESY6,Abbott2022} and KiDS \citep{Heymans2021}. This 
combination is powerful because it enables self-calibration of galaxy 
bias and provides complementary constraints on cosmological parameters. 
The galaxy-galaxy lensing signal depends on the product of galaxy bias 
and the matter amplitude, breaking degeneracies present in either probe 
alone.

Recent analyses combining galaxy-galaxy lensing with spectroscopic 
surveys have yielded particularly interesting results. 
\citet{luo2025} find $S_8 = 0.8294 \pm 0.0110$ from HSC lensing 
combined with BOSS spectroscopy, in excellent agreement with the CMB 
baseline and suggesting that the low $S_8$ values from some cosmic shear 
analyses may reflect systematic effects rather than new physics.

\subsection{Galaxy Cluster Counts}

The abundance of galaxy clusters as a function of mass and redshift provides a sensitive probe of the growth of structure. The halo mass function $dn/dM$ depends exponentially on $\sigma_8$ at the high-mass end, making cluster counts particularly sensitive to the clustering amplitude. The observable-mass relation (connecting X-ray luminosity, SZ signal, or richness to true mass) represents the primary systematic uncertainty in cluster cosmology. The mass-richness relation (for optical clusters) or hydrostatic bias (for X-ray/SZ) are the specific dominant nuisance parameters driving the uncertainties.

Multiple cluster analyses have contributed to the $S_8$ landscape in recent years. \citet{Ghirardini2024} present X-ray cluster cosmology from the eROSITA All-Sky Survey, finding $S_8 = 0.86 \pm 0.01$---notably higher than other late-universe probes, though the discrepancy stands at only $1.5\sigma$. In contrast, SPT cluster analyses \citep{Bocquet2024} favor lower values ($S_8 \approx 0.79$--$0.80$), creating an internal tension within cluster cosmology that remains unresolved. Earlier analyses combining \textit{Planck} SZ clusters with various mass calibration 
approaches have yielded $S_8$ values consistent with or lower than the SPT results, 
rather than intermediate between SPT and eROSITA. \citet{Zubeldia2019}, using CMB 
lensing mass calibration, report $\sigma_8 = 0.76 \pm 0.04$ and 
$\Omega_\mathrm{m} = 0.33 \pm 0.02$, corresponding to 
$S_8 \approx 0.80 \pm 0.05$\footnote{\label{fn:zubeldia}Derived from the reported $\sigma_8$ and
$\Omega_\mathrm{m}$ values using the standard definition 
$S_8 \equiv \sigma_8 \sqrt{\Omega_\mathrm{m}/0.3}$. The original paper reports the 
combination $\sigma_8 (\Omega_\mathrm{m}/0.33)^{0.25} = 0.765 \pm 0.035$ as its 
best-constrained parameter.}. \citet{Aymerich2024}, using Chandra X-ray and CFHT 
weak-lensing mass calibration, find $S_8 = 0.78 \pm 0.02$ for their baseline 
self-similar redshift evolution model and $S_8 = 0.81 \pm 0.02$ when the redshift 
evolution is left free. These values cluster around or below the SPT result 
($S_8 \approx 0.80$), reinforcing a pattern of low $S_8$ from SZ cluster number 
counts and highlighting the discrepancy with the eROSITA value of 
$S_8 = 0.86 \pm 0.01$.

\subsection{Redshift-Space Distortions}

The peculiar velocities of galaxies induce anisotropies in the observed clustering 
pattern when distances are inferred from redshifts. These redshift-space distortions 
(RSD) provide a direct measurement of the growth rate $f\sigma_8$, where 
$f \equiv d\ln D/d\ln a$ is the logarithmic growth rate and $D$ is the linear growth 
factor.

RSD measurements have been extensively used to probe the $S_8$ tension. 
\citet{Nunes2021} combined RSD data with geometrical measurements from BAO and 
Type~Ia supernovae to arbitrate between early- and late-universe determinations of 
$S_8$, finding $S_8 = 0.762^{+0.030}_{-0.025}$ and concluding that there are hints, 
but no strong evidence yet, for suppressed growth relative to \textit{Planck}. 
\citet{Benisty2021} quantified the tension using RSD data alone, finding 
$S_8 = 0.700^{+0.038}_{-0.037}$ from Bayesian analysis, corresponding to a $3.4\sigma$ 
discrepancy with \textit{Planck} ($3.5\sigma$ relative to the Combined CMB baseline adopted in this review), while a model-independent Gaussian Process 
reconstruction yielded $S_8 \approx 0.70$--$0.73$ with the tension reduced to 
$\sim 1.5\sigma$ due to the larger uncertainties of the non-parametric approach. 
More recent analyses incorporating DESI full-shape clustering measurements, which 
include RSD information and constrain $\sigma_8 = 0.842 \pm 0.034$ 
\citep{DESI2024FullShape}, and joint RSD+CMB fits \citep{Sabogal2024} continue to 
probe the consistency of growth rates with $\Lambda$CDM predictions.

The studies by \citet{Adil2024} and \citet{Akarsu2024} present an intriguing 
finding: $S_8$ appears to increase with effective redshift when inferred from 
different probes, potentially indicating scale-dependent or time-dependent 
modifications to structure formation within $\Lambda$CDM cosmology.

%==============================================================================
\section{Compilation of \texorpdfstring{$S_8$}{S8} Measurements (2019--2026)}
%==============================================================================

\subsection{Visual Summary}

\begin{figure}[ht!]
    \centering
    \begin{tikzpicture}
    \begin{axis}[
        width=0.92\textwidth,
        height=13cm,
        xlabel={Year of Publication},
        ylabel={$S_8 \equiv \sigma_8(\Omega_m/0.3)^{0.5}$},
        xmin=2019.0, xmax=2026.7,
        ymin=0.65, ymax=0.90,
        x tick label style={/pgf/number format/1000 sep=},
        xtick={2020, 2021, 2022, 2023, 2024, 2025, 2026},
        ytick={0.65, 0.68, 0.70, 0.72, 0.74, 0.76, 0.78, 0.80, 0.82, 0.84, 0.86, 0.88, 0.90},
        legend style={
            at={(0.98,0.02)},
            anchor=south east,
            font=\footnotesize,
            fill=white,
            fill opacity=0.95,
            draw=gray,
            cells={anchor=west},
            row sep=1pt,
            /tikz/every even column/.append style={column sep=5pt}
        },
        legend columns=1,
        grid=major,
        grid style={dashed, gray!30},
        clip=false
    ]

    % Combined CMB Baseline (Gray Band) - asymmetric: 0.836 +0.012 / -0.013
    \addplot [name path=upper, draw=none, forget plot] coordinates {(2019.3, 0.848) (2026.7, 0.848)};
    \addplot [name path=lower, draw=none, forget plot] coordinates {(2019.3, 0.823) (2026.7, 0.823)};
    \addplot [gray!25, forget plot] fill between [of=upper and lower];
    \addplot [gray, thick, dashed, forget plot] coordinates {(2019.3, 0.836) (2026.7, 0.836)};
    \node[gray, anchor=south west, font=\footnotesize\bfseries, fill=white, fill opacity=0.8, inner sep=2pt] at (axis cs: 2024.5, 0.852) {Combined CMB $1\sigma$};
    
    % --- DATA POINTS (asymmetric error bars via table format) ---
    
    % CMB Points (Black squares) — all symmetric
    \addplot[only marks, mark=square*, mark size=3pt, color=black, 
        error bars/.cd, y dir=both, y explicit]
    table [x index=0, y index=1, y error plus index=2, y error minus index=3] {
        2020.0  0.832  0.013  0.013
        2025.8  0.875  0.023  0.023
        2026.0  0.828  0.012  0.012
    };
    \addlegendentry{CMB \& CMB Lensing}

    % DES Data (Red circles) — asymmetric errors
    \addplot[only marks, mark=*, mark size=3.5pt, color=red!85!black, 
        error bars/.cd, y dir=both, y explicit]
    table [x index=0, y index=1, y error plus index=2, y error minus index=3] {
        2022.1  0.776  0.017  0.017
        2024.2  0.823  0.019  0.020
        2026.0  0.794  0.009  0.012
        2026.1  0.789  0.012  0.012
        2026.2  0.783  0.019  0.015
        2026.3  0.781  0.021  0.020
        2026.5  0.747  0.027  0.025
    };
    \addlegendentry{Dark Energy Survey (DES)}

    % KiDS Data (Blue circles) — asymmetric errors
    \addplot[only marks, mark=*, mark size=3.5pt, color=blue!85!black, 
        error bars/.cd, y dir=both, y explicit]
    table [x index=0, y index=1, y error plus index=2, y error minus index=3] {
        2021.0  0.759  0.024  0.021
        2023.8  0.790  0.018  0.014
        2025.2  0.815  0.016  0.021
    };
    \addlegendentry{Kilo-Degree Survey (KiDS)}

    % Cluster Counts (Orange triangles) — all symmetric
    \addplot[only marks, mark=triangle*, mark size=4pt, color=orange!90!black, 
        error bars/.cd, y dir=both, y explicit]
    table [x index=0, y index=1, y error plus index=2, y error minus index=3] {
        2019.5  0.800  0.050  0.050
        2024.0  0.795  0.029  0.029
        2024.1  0.780  0.020  0.020
        2024.7  0.860  0.010  0.010
    };
    \addlegendentry{Galaxy Cluster Counts}

    % HSC (Green diamonds) — asymmetric errors
    \addplot[only marks, mark=diamond*, mark size=4pt, color=green!65!black, 
        error bars/.cd, y dir=both, y explicit]
    table [x index=0, y index=1, y error plus index=2, y error minus index=3] {
        2019.2  0.780  0.030  0.033
        2023.4  0.812  0.021  0.021
        2023.6  0.775  0.043  0.038
        2023.9  0.776  0.032  0.033
        2025.4  0.805  0.018  0.018
    };
    \addlegendentry{Hyper Suprime-Cam (HSC)}

    % Other probes (Purple pentagons) — some asymmetric
    \addplot[only marks, mark=pentagon*, mark size=3.5pt, color=purple!80!black, 
        error bars/.cd, y dir=both, y explicit]
    table [x index=0, y index=1, y error plus index=2, y error minus index=3] {
        2020.9  0.776  0.033  0.033
        2021.2  0.762  0.030  0.025
        2021.5  0.700  0.038  0.037
        2023.0  0.792  0.022  0.022
        2024.3  0.816  0.024  0.024
        2024.5  0.775  0.027  0.027
        2025.0  0.808  0.017  0.017
        2025.1  0.829  0.011  0.011
        2025.5  0.819  0.030  0.030
        2025.9  0.810  0.050  0.050
    };
    \addlegendentry{Other (DESI, RSD, GGL, Ly$\alpha$, Sat.\ Kin.)}

    % Key annotations
    \node[red!85!black, font=\scriptsize, anchor=north west] at (axis cs: 2026.2, 0.785) {DES Y6};
    \node[red!85!black, font=\scriptsize, anchor=north] at (axis cs: 2026.5, 0.740) {DES Y1};
    \node[red!85!black, font=\scriptsize, anchor=south east] at (axis cs: 2024.1, 0.825) {WL+kSZ};
    \node[blue!85!black, font=\scriptsize, anchor=south west] at (axis cs: 2025.3, 0.817) {KiDS};
    \node[orange!90!black, font=\scriptsize, anchor=south] at (axis cs: 2024.7, 0.875) {eROSITA};
    \node[black, font=\scriptsize, anchor=east] at (axis cs: 2019.9, 0.832) {Planck};
    \node[black, font=\scriptsize, anchor=west] at (axis cs: 2025.9, 0.880) {ACT DR6};
    \node[purple!80!black, font=\scriptsize, anchor=east] at (axis cs: 2021.4, 0.693) {Benisty};
    \node[purple!80!black, font=\scriptsize, anchor=east] at (axis cs: 2021.1, 0.755) {Nunes};

    \end{axis}
    \end{tikzpicture}
    \caption{Compilation of $S_8$ measurements from major cosmological surveys (2019--2026). The horizontal gray band represents the $1\sigma$ confidence interval of the Combined CMB baseline (Planck + ACT DR6 + SPT-3G), centered at $S_8 = 0.836$ with asymmetric width $^{+0.012}_{-0.013}$. The dashed gray line indicates the central value. Note the systematic offset between DES results (red circles), which lie consistently below the CMB band, and KiDS Legacy results (blue circles), which have migrated upward into consistency with the CMB. The eROSITA cluster count measurement (orange triangle, upper region) uniquely favors a clustering amplitude exceeding the CMB baseline. Error bars represent $1\sigma$ uncertainties as reported; asymmetric uncertainties are shown faithfully.}
    \label{fig:s8_plot}
\end{figure}

Several features are immediately apparent from this visualization. The DES measurements (red points) lie systematically below the CMB band across all data releases, with the Y6 results representing the most precise low-$S_8$ determination to date. In contrast, the KiDS results (blue points) show significant evolution: the KiDS-1000 result from 2021 \citep{Asgari2021} exhibited a pronounced tension similar to DES, but the KiDS Legacy result from 2025 \citep{Wright2025} has shifted upward into consistency with the CMB. The eROSITA cluster count measurement \citep{Ghirardini2024} stands out as the sole late-universe probe favoring $S_8$ values above the CMB baseline. 

The spectroscopic and RSD probes (purple pentagons) span a notably wide range, from 
$S_8 = 0.700$ \citep{Benisty2021} to $S_8 = 0.829$ \citep{luo2025}, reflecting the 
diversity of methodologies within this category. Galaxy cluster counts (orange triangles) 
exhibit striking internal disagreement, with eROSITA favouring $S_8$ above the CMB band 
while SPT and Planck SZ clusters fall well below. Galaxy-galaxy lensing and peculiar 
velocity measurements cluster near the CMB baseline, suggesting consistency when weak 
lensing systematics are avoided.

\subsection{Detailed Tabulation}

Table \ref{tab:s8_measurements} provides comprehensive information for each measurement, including the probe type, publication date, $S_8$ value with uncertainties, tension level relative to the Combined CMB baseline, and relevant citations.

\begin{longtable}{p{3.2cm} c c c p{5.0cm}}
\caption{Comprehensive compilation of $S_8$ measurements and their tension with the Combined CMB baseline ($S_8 = 0.836^{+0.012}_{-0.013}$). Tensions are computed using the uncertainty in the direction of the discrepancy for asymmetric errors. Negative values indicate $S_8$ above the baseline.} \label{tab:s8_measurements} \\
\toprule
\textbf{Probe / Analysis} & \textbf{Date} & \textbf{$S_8$ Value} & \textbf{Tension} & \textbf{Reference \& Notes} \\
\midrule
\endfirsthead
\toprule
\textbf{Probe / Analysis} & \textbf{Date} & \textbf{$S_8$ Value} & \textbf{Tension} & \textbf{Reference \& Notes} \\
\midrule
\endhead
\midrule
\multicolumn{5}{r}{\textit{Continued on next page...}} \\
\endfoot
\bottomrule
\endlastfoot

\multicolumn{5}{l}{\textbf{CMB \& Early Universe Baselines}} \\
\midrule
Combined CMB & Jan 2026 & $0.836^{+0.012}_{-0.013}$ & --- & Planck+ACT+SPT; primary baseline \citep{DESY6} \\
\addlinespace
Combined Lensing & Jan 2026 & $0.828 \pm 0.012$ & $0.5\sigma$ & ACT+SPT+\textit{Planck} lensing + DESI DR2 BAO \citep{Qu2026} \\
\addlinespace
ACT DR6 (primary CMB) & Nov 2025 & $0.875 \pm 0.023$ & $-1.5\sigma$ & ACT DR6 power spectra alone (+ Sroll2 $\tau$ prior); above baseline \citep{Louis2025} \\
\addlinespace
Planck 2018 & Sep 2020 & $0.832 \pm 0.013$ & --- & Previous baseline \citep{Planck2020} \\
\midrule

\multicolumn{5}{l}{\textbf{Dark Energy Survey}} \\
\midrule
DES Y6 (All) & Jan 2026 & $0.794^{+0.009}_{-0.012}$ & $2.4\sigma$ & $3\times2\mathrm{pt}$ + BAO + SNe + clusters \citep{DESY6} \\
\addlinespace
DES Y6 ($3\times2\mathrm{pt}$) & Jan 2026 & $0.789 \pm 0.012$ & $2.7\sigma$ & $3\times2\mathrm{pt}$ analysis (cosmic shear + galaxy clustering + galaxy-galaxy lensing) \citep{DESY6} \\
\addlinespace
DES Y6 (Shear) & Jan 2026 & $0.783^{+0.019}_{-0.015}$ & $2.7\sigma$ & Cosmic shear only \citep{DESY6} \\
\addlinespace
DES Y6 ($w$CDM) & Jan 2026 & $0.781^{+0.021}_{-0.020}$ & $1.0\sigma$ & Extended model; reported tension (projected) \citep{DESY6} \\
\addlinespace
DES Y3 WL + ACT kSZ & Apr 2024 & $0.823^{+0.019}_{-0.020}$ & $0.6\sigma$ & Joint WL + kSZ with BCEmu7 baryonification \citep{Bigwood2024} \\
\addlinespace
DES Y3 ($3\times2\mathrm{pt}$) & Jan 2022 & $0.776 \pm 0.017$ & $2.8\sigma$ & Previous DES release \citep{Abbott2022} \\
\addlinespace
DES Y1 ($3\times2\mathrm{pt}$) & Jan 2026 & $0.747^{+0.027}_{-0.025}$ & $3.2\sigma$ & Reanalysis with Y6 pipeline \citep{DESY6} \\
\midrule

\multicolumn{5}{l}{\textbf{Kilo-Degree Survey}} \\
\midrule
KiDS-Legacy (Shear) & Mar 2025 & $0.815^{+0.016}_{-0.021}$ & $0.9\sigma$ & Final KiDS result \citep{Wright2025} \\
\addlinespace
KiDS-1000 + DES Y3 & Oct 2023 & $0.790^{+0.018}_{-0.014}$ & $2.4\sigma$ & Joint cosmic shear analysis \citep{Kilo-DegreeSurvey:2023gfr} \\
\addlinespace
KiDS-1000 (Shear) & Jan 2021 & $0.759^{+0.024}_{-0.021}$ & $\mathbf{3.1\sigma}$ & Superseded \citep{Asgari2021} \\
\midrule

\multicolumn{5}{l}{\textbf{Hyper Suprime-Cam}} \\
\midrule
HSC Y3 (DESI n(z))    & Nov 2025 & $0.805 \pm 0.018$ & $1.4\sigma$ & Reanalysis with DESI clustering-z calibration \citep{ChoppinDeJanvry2025}\\
\addlinespace
HSC Y3 ($3\times 2$pt, minimal bias) & Dec 2023 & $0.775^{+0.043}_{-0.038}$ & $1.5\sigma$ & HSC shear + SDSS clustering \citep{Sugiyama2023} \\
\addlinespace
HSC Y3 (Shear) & Dec 2023 & $0.776^{+0.032}_{-0.033}$ & $1.7\sigma$ & Third-year analysis \citep{Dalal2023} \\
\addlinespace
HSC Y1 (Unified) & Apr 2023 & $0.812 \pm 0.021$ & $1.0\sigma$ & Unified pipeline \citep{Longley2023} \\
\addlinespace
HSC Y1 (Original) & Mar 2019 & $0.780^{+0.030}_{-0.033}$ & $1.6\sigma$ & Original analysis \citep{Hikage2019} \\
\midrule

\multicolumn{5}{l}{\textbf{Galaxy Cluster Counts}} \\
\midrule
eROSITA (X-ray) & Oct 2024 & $0.86 \pm 0.01$ & $-1.5\sigma$ & Higher than CMB \citep{Ghirardini2024} \\
\addlinespace
SPT-3G Clusters & Jan 2024 & $0.795 \pm 0.029$ & $1.3\sigma$ & SZ-selected \citep{Bocquet2024} \\
\addlinespace
\textit{Planck} SZ (Chandra+WL) & Feb 2024 & $0.78 \pm 0.02$ & $2.3\sigma$ & PSZ2 cluster counts, Chandra X-ray + CFHT WL mass calibration; baseline self-similar model \citep{Aymerich2024} \\
\addlinespace
\textit{Planck}+CMB lensing & 2019 & $0.80 \pm 0.05$\textsuperscript{\ref{fn:zubeldia}} & $0.7\sigma$ & Lensing mass calibration \citep{Zubeldia2019} \\
\midrule

\multicolumn{5}{l}{\textbf{Spectroscopic \& Other Probes}} \\
\midrule
Satellite Kinematics (SDSS) & Dec 2025 & $0.81 \pm 0.05$ & $0.5\sigma$ & Basilisk: sat.\ kinematics + luminosity function \citep{Mitra2025} \\
GGL (HSC$\times$BOSS) & Feb 2025 & $0.8294 \pm 0.0110$ & $0.4\sigma$ & Galaxy-galaxy lensing \citep{luo2025} \\
\addlinespace
DESI DR1 + Lensing & May 2025 & $0.808 \pm 0.017$ & $1.3\sigma$ & 3D clustering + CMB lensing \citep{Maus2025} \\
\addlinespace
Peculiar Velocities & Sep 2025 & $0.819 \pm 0.030$ & $0.5\sigma$ & TFR+FP+SNe joint \citep{Stiskalek2025} \\
\addlinespace
BOSS$\times$DECaLS (Magnification) & Sep 2024 & $0.816 \pm 0.024$ & $0.7\sigma$ & Lensing magnification of CMASS galaxies~\citep{Xu2024}\\
\addlinespace
RSD compilation & Aug 2024 & $0.775 \pm 0.027$ & $2.0\sigma$ & RSD+DESI+PP \citep{Sabogal2024} \\
\addlinespace
BOSS Full-Shape & 2023 & $0.792 \pm 0.022$ & $1.7\sigma$ & Clustering + GGL \citep{Lange2023BOSS} \\
\addlinespace
RSD (Benisty) & 2021 & $0.700^{+0.038}_{-0.037}$ & $3.5\sigma$ & RSD compilation \citep{Benisty2021} \\
\addlinespace
RSD+BAO+SNe & 2021 & $0.762^{+0.030}_{-0.025}$ & $2.6\sigma$ & RSD compilation + BAO + SNeIa \citep{Nunes2021} \\
\addlinespace
Ly$\alpha$ Forest & Nov 2020 & $0.776 \pm 0.033$\footnotemark & $1.7\sigma$ & SDSS DR14 \citep{Palanque2020}\\
\end{longtable}

\footnotetext{Derived from the reported $\sigma_8 = 0.811 \pm 0.024$ and $\Omega_m = 0.275 \pm 0.012$ using $S_8 \equiv \sigma_8\sqrt{\Omega_m/0.3}$.}

\subsection{Temporal Evolution}

\begin{figure}[ht!]
    \centering
    \begin{tikzpicture}
    \begin{axis}[
        width=0.92\textwidth,
        height=10cm,
        xlabel={Year of Publication},
        ylabel={$S_8 \equiv \sigma_8(\Omega_m/0.3)^{0.5}$},
        xmin=2018.5, xmax=2026.8,
        ymin=0.70, ymax=0.88,
        x tick label style={/pgf/number format/1000 sep=},
        xtick={2019, 2020, 2021, 2022, 2023, 2024, 2025, 2026},
        ytick={0.70, 0.72, 0.74, 0.76, 0.78, 0.80, 0.82, 0.84, 0.86, 0.88},
        grid=major,
        grid style={dashed, gray!30},
        legend style={
            at={(0.98,0.02)},
anchor=south east,
            font=\footnotesize,
            fill=white,
            fill opacity=0.95,
            draw=gray,
            cells={anchor=west},
            row sep=1pt,
        },
        clip=false
    ]

    % =============================================
    % Combined CMB Baseline (Red Band): 0.836 +0.012 / -0.013
    % =============================================
    \addplot [name path=cmb_upper, draw=none, forget plot] 
        coordinates {(2018.5, 0.848) (2026.8, 0.848)};
    \addplot [name path=cmb_lower, draw=none, forget plot] 
        coordinates {(2018.5, 0.823) (2026.8, 0.823)};
    \addplot [red!15, forget plot] fill between [of=cmb_upper and cmb_lower];
    \addplot [red!60, thick, dashed, forget plot] 
        coordinates {(2018.5, 0.836) (2026.8, 0.836)};

    % Band labels
    \node[red!70!black, anchor=south west, font=\footnotesize\bfseries, 
          fill=white, fill opacity=0.8, inner sep=2pt] 
        at (axis cs: 2018.7, 0.849) {Combined CMB: $S_8 = 0.836^{+0.012}_{-0.013}$};

    % =============================================
    % Individual measurements (small gray dots)
    % =============================================
    \addplot[only marks, mark=*, mark size=1.5pt, color=gray!50, 
        error bars/.cd, y dir=both, y explicit, 
        error bar style={line width=0.4pt, gray!40}]
    coordinates {
        % 2019
        (2019.2, 0.780) +- (0.0, 0.031)
        (2019.5, 0.800) +- (0.0, 0.050)
        % 2020
        (2020.0, 0.776) +- (0.0, 0.033)
        % 2021
        (2021.0, 0.759) +- (0.0, 0.023)
        (2021.2, 0.762) +- (0.0, 0.028)
        (2021.5, 0.700) +- (0.0, 0.038)
        % 2022
        (2022.1, 0.776) +- (0.0, 0.017)
        % 2023
        (2023.0, 0.792) +- (0.0, 0.022)
        (2023.3, 0.812) +- (0.0, 0.021)
        (2023.5, 0.775) +- (0.0, 0.041)
        (2023.7, 0.790) +- (0.0, 0.016)
        (2023.9, 0.776) +- (0.0, 0.033)
        % 2024
        (2024.0, 0.795) +- (0.0, 0.029)
        (2024.1, 0.780) +- (0.0, 0.020)
        (2024.3, 0.816) +- (0.0, 0.024)
        (2024.5, 0.775) +- (0.0, 0.027)
        (2024.7, 0.860) +- (0.0, 0.010)
        (2024.2, 0.823) +- (0.0, 0.020)
        % 2025
        (2025.0, 0.808) +- (0.0, 0.017)
        (2025.1, 0.829) +- (0.0, 0.011)
        (2025.2, 0.815) +- (0.0, 0.019)
        (2025.4, 0.805) +- (0.0, 0.018)
        (2025.5, 0.819) +- (0.0, 0.030)
        (2025.9, 0.810) +- (0.0, 0.050)
        % 2026
        (2026.0, 0.794) +- (0.0, 0.011)
        (2026.1, 0.789) +- (0.0, 0.012)
        (2026.2, 0.783) +- (0.0, 0.017)
        (2026.3, 0.781) +- (0.0, 0.021)
        (2026.5, 0.747) +- (0.0, 0.026)
    };
    \addlegendentry{Individual measurements}

    % =============================================
    % Dashed line connecting weighted means
    % =============================================
    \addplot[densely dashed, color=blue!70!black, line width=0.8pt, forget plot]
    coordinates {
        (2019, 0.7857)
        (2020, 0.7760)
        (2021, 0.7495)
        (2022, 0.7760)
        (2023, 0.7933)
        (2024, 0.8309)
        (2025, 0.8185)
        (2026, 0.7865)
    };

    % =============================================
    % Weighted mean points (blue squares)
    % =============================================
    \addplot[only marks, mark=square*, mark size=3.5pt, color=blue!70!black,
        mark options={fill=blue!70!black},
        error bars/.cd, y dir=both, y explicit, 
        error bar style={line width=1.2pt, blue!70!black},
        error mark options={rotate=90, mark size=3pt, line width=0.8pt, blue!70!black}]
    coordinates {
        (2019, 0.7857) +- (0.0, 0.0267)   % N=2, 1.7σ
        (2020, 0.7760) +- (0.0, 0.0330)   % N=1, 1.7σ
        (2021, 0.7495) +- (0.0, 0.0158)   % N=3, 4.2σ
        (2022, 0.7760) +- (0.0, 0.0170)   % N=1, 2.8σ
        (2023, 0.7933) +- (0.0, 0.0101)   % N=5, 2.6σ
        (2024, 0.8309) +- (0.0, 0.0072)   % N=6, 0.3σ
        (2025, 0.8185) +- (0.0, 0.0072)   % N=6, 1.2σ
        (2026, 0.7865) +- (0.0, 0.0066)   % N=5, 3.4σ
    };
    \addlegendentry{Weighted mean (per year)}

    % ============================================= 
    % N annotations (below points)
    % =============================================
    \node[font=\scriptsize, color=blue!60!black] at (axis cs: 2019, 0.751) {$N\!=\!2$};
    \node[font=\scriptsize, color=blue!60!black] at (axis cs: 2020, 0.735) {$N\!=\!1$};
    \node[font=\scriptsize, color=blue!60!black] at (axis cs: 2021, 0.726) {$N\!=\!3$};
    \node[font=\scriptsize, color=blue!60!black] at (axis cs: 2022, 0.751) {$N\!=\!1$};
    \node[font=\scriptsize, color=blue!60!black] at (axis cs: 2023, 0.775) {$N\!=\!5$};
    \node[font=\scriptsize, color=blue!60!black] at (axis cs: 2024, 0.818) {$N\!=\!6$};
    \node[font=\scriptsize, color=blue!60!black] at (axis cs: 2025, 0.803) {$N\!=\!6$};
    \node[font=\scriptsize, color=blue!60!black] at (axis cs: 2026, 0.772) {$N\!=\!5$};

    % =============================================
    % Tension annotations (above points)
    % =============================================
    \node[font=\scriptsize, color=black!70, anchor=south] at (axis cs: 2019, 0.815) {$1.7\sigma$};
    \node[font=\scriptsize, color=black!70, anchor=south] at (axis cs: 2020, 0.812) {$1.7\sigma$};
    \node[font=\scriptsize, color=black!70, anchor=south] at (axis cs: 2021, 0.768) {$4.2\sigma$};
    \node[font=\scriptsize, color=black!70, anchor=south] at (axis cs: 2022, 0.796) {$2.8\sigma$};
    \node[font=\scriptsize, color=black!70, anchor=south] at (axis cs: 2023, 0.806) {$2.6\sigma$};
    \node[font=\scriptsize, color=black!70, anchor=south west] at (axis cs: 2024.1, 0.840) {$0.3\sigma$};
    \node[font=\scriptsize, color=black!70, anchor=south] at (axis cs: 2025, 0.829) {$1.2\sigma$};
    \node[font=\scriptsize, color=black!70, anchor=south] at (axis cs: 2026, 0.796) {$3.4\sigma$};

    \end{axis}
    \end{tikzpicture}
    \caption{Weighted mean $S_8$ per year of publication (blue squares) computed from 
    all late-universe measurements in Table~\ref{tab:s8_measurements}, excluding 
    CMB baselines. Individual measurements are shown as gray points. The red band 
    denotes the $1\sigma$ interval of the Combined CMB baseline 
    ($S_8 = 0.836^{+0.012}_{-0.013}$). For asymmetric uncertainties, the 
    symmetrized error $\sigma_{\rm sym} = \sqrt{(\sigma_+^2 + \sigma_-^2)/2}$ is used 
    to compute inverse-variance weights. The number of measurements per year is 
    indicated below each point, and the tension with the Combined CMB baseline 
    above. The high 2024 mean ($0.3\sigma$) is driven by the eROSITA cluster 
    count result ($S_8 = 0.86 \pm 0.01$), which dominates the inverse-variance 
    weighting; the 2026 mean is dominated by the DES~Y6 results. 
    Correlations between measurements sharing underlying data are not 
    accounted for.}
    \label{fig:s8_weighted_mean}
\end{figure}
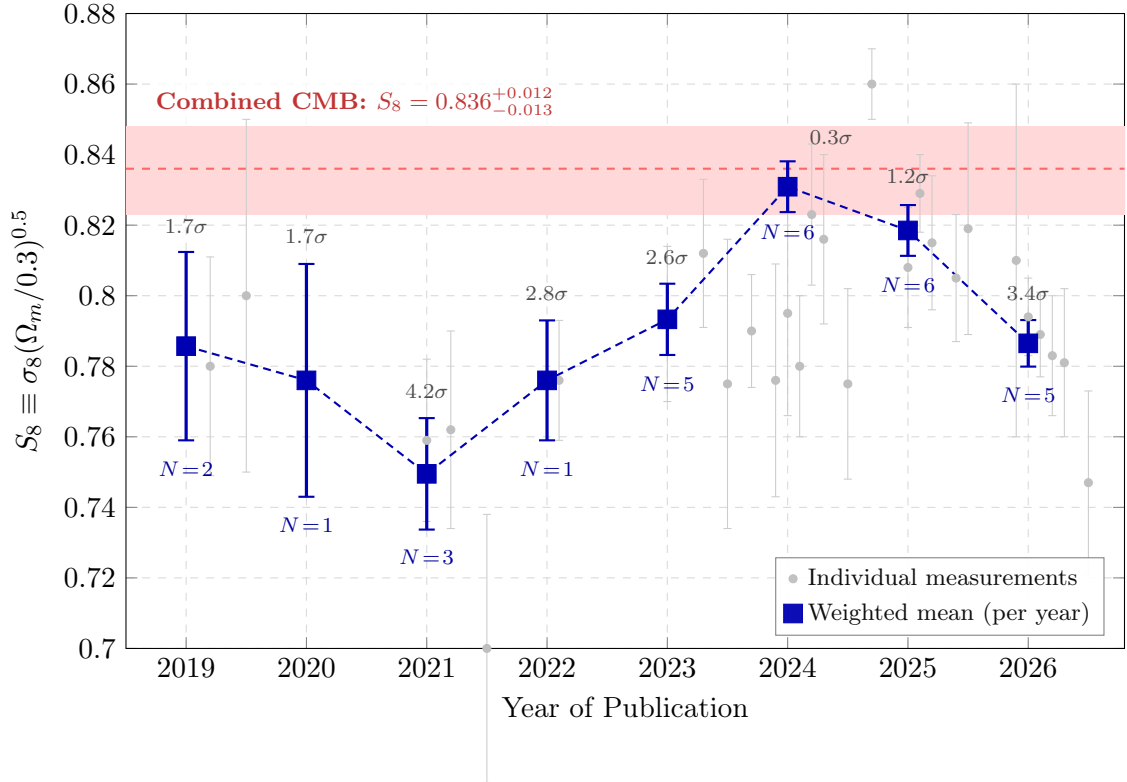

Figure~\ref{fig:s8_weighted_mean} presents an illustrative inverse-variance weighted mean $S_8$ computed annually from all late-universe measurements in  Table~\ref{tab:s8_measurements}, excluding CMB baselines. For asymmetric uncertainties, we adopt the symmetrized error $\sigma_{\rm sym} = \sqrt{(\sigma_+^2 + \sigma_-^2)/2}$.

The overall pattern reveals a broad convergence toward the 
Combined CMB baseline over the period 2019--2025, accompanied 
by a steady reduction in the annual uncertainties as the number 
and precision of measurements have increased. The early years 
(2019--2022) are characterized by sparse data and representative weighted means 
in the range $S_8 \approx 0.75$--$0.79$, with tensions of 
$1.7\sigma$--$4.2\sigma$. The 2021 value ($S_8 = 0.750 \pm 0.016$, 
$4.2\sigma$) represents the peak of the reported tension; however, 
this is substantially influenced by the \citet{Benisty2021} RSD 
result ($S_8 = 0.700$), which lies well below all other 
measurements in our compilation. The composite weighted mean rises through 
2023 ($S_8 = 0.793 \pm 0.010$, $2.6\sigma$) and reaches 
consistency with the CMB in 2024 ($S_8 = 0.831 \pm 0.007$, 
$0.3\sigma$), before dropping sharply in 2026 ($S_8 = 0.787 \pm 0.007$, $3.4\sigma$).

The 2024 indicative weighted mean requires careful interpretation. The 
eROSITA cluster count measurement ($S_8 = 0.86 \pm 0.01$; 
\citealt{Ghirardini2024}), with its small uncertainty, 
dominates the inverse-variance weighting and pulls the annual 
mean above the CMB baseline. Excluding eROSITA, the 2024
weighted mean drops to $S_8 = 0.799 \pm 0.010$ ($2.2\sigma$ 
tension), restoring consistency with the overall low-$S_8$ trend 
from late-universe probes. This underscores the anomalous 
character of the eROSITA result relative to other probes in the 
same epoch.

The 2025 annual  indicative mean ($S_8 = 0.819 \pm 0.007$, $1.2\sigma$) 
is notable for representing the most methodologically diverse 
year in the compilation: six measurements spanning weak lensing 
(KiDS Legacy, HSC~Y3 recalibration), galaxy-galaxy lensing 
(HSC$\times$BOSS), spectroscopic clustering (DESI~DR1), peculiar 
velocities, and satellite kinematics (Basilisk). Despite this 
diversity, all six measurements are mutually consistent and 
cluster within the range $S_8 = 0.805$--$0.829$, suggesting 
a genuine convergence across independent methodologies toward 
a value moderately below the CMB prediction.

The 2026 value is dominated by the DES~Y6 results, 
which constitute all five measurements published in that year. 
The DES~Y6 combined analysis ($S_8 = 0.794^{+0.009}_{-0.012}$), 
with the smallest uncertainty of any individual measurement in 
our compilation, anchors the weighted mean. The inclusion of the 
DES~Y1 reanalysis with the Y6 pipeline ($S_8 = 0.747$; 
\citealt{DESY6}), which yields a notably low value, further 
pulls the mean downward; excluding it shifts the weighted mean 
from $S_8 = 0.787$ to $0.789$ ($3.2\sigma$), a modest change 
that confirms the dominance of the Y6 combined result.

We caution that this visualization does not account for 
correlations between measurements that share underlying data 
(e.g., the multiple DES~Y6 results in 2026, or KiDS-1000 and 
the KiDS$+$DES joint analysis in 2021 and 2023). The annual 
weighted means should therefore be interpreted as descriptive 
summaries of the evolving measurement landscape rather than 
statistically rigorous combined constraints.

It is important to emphasize that the annual weighted means provided in Figure~\ref{fig:s8_weighted_mean} are intended to be illustrative of the general trend rather than a rigorous meta-analysis. Since several measurements share underlying data or systematic pipelines, a full covariance matrix would be required for a formal combination. Consequently, the reported $n_{\sigma}$ tension levels in this section should be interpreted as descriptive indicators of the evolving landscape

%==============================================================================
\section{Analysis of Key Results}
%==============================================================================

\subsection{The Dark Energy Survey Year 6 Analysis}

The DES Year 6 (Y6) analysis \citep{DESY6} represents the culmination of the Dark Energy Survey's imaging program, utilizing the full six years of observations collected with the Dark Energy Camera on the 4-meter Blanco telescope at Cerro Tololo Inter-American Observatory. The Y6 dataset encompasses approximately 5000 deg$^2$ of the southern sky, reaching limiting magnitudes of $g \approx 24.7$, $r \approx 24.4$, $i \approx 23.8$, and $z \approx 23.0$ (10$\sigma$ point source).

\subsubsection{Methodological Advances}

The Y6 analysis incorporates several methodological improvements relative to the Y3 release \citep{Abbott2022}. The shear catalog employs the METADETECTION algorithm \citep{Yamamoto:2025jsd}, an evolution of METACALIBRATION that uses image manipulations to mitigate detection and shear measurement biases. However, the analysis explicitly utilizes a suite of image simulations to characterize galaxy blending and determine priors for the multiplicative shear bias. This approach achieves multiplicative bias uncertainties ranging from approximately 0.6\% to 1.2\%.

The photometric redshift distributions are calibrated using a hybrid framework that combines the Self-Organizing Map $n(z)$ method (SOMPZ), which maps spectroscopic information from deep fields to the wide survey using transfer functions from image simulations, with clustering-redshift cross-correlations (WZ) against spectroscopic samples \citep{DESY6}. The intrinsic alignment model employs the tidal alignment and tidal torquing (TATT) framework, marginalizing over amplitude and redshift-evolution parameters ($A_1, A_2, \eta_1, \eta_2$) using informative priors on the evolution terms motivated by observations, while holding the tidal alignment bias parameter $b_{\text{TA}}$ fixed \citep{DESY6}.

\subsubsection{Results and Tension Assessment}

The DES Y6 3$\times$2pt analysis yields $S_8 = 0.789 \pm 0.012$, representing a $2.7\sigma$ tension with the Combined CMB baseline in the $S_8$ projection. This tension remains consistent when comparing to the cosmic shear signal alone, which favors an even lower value of $S_8 = 0.783^{+0.019}_{-0.015}$, suggesting that the tension is not an artifact of galaxy bias modeling in the clustering and galaxy-galaxy lensing components.

The combination of DES Y6 3$\times$2pt with other internal DES probes (DES BAO, DES Supernovae, and DES Cluster Counts) yields the tightest internal constraint, $S_8 = 0.794^{+0.009}_{-0.012}$. This combined result exhibits a $2.8\sigma$ tension with the Combined CMB in the full parameter space ($2.4\sigma$ when projected onto $S_8$). Notably, an independent reanalysis of the DES Y3 $3\times 2$pt data using one-loop predictions from the Effective Field Theory of Large-Scale Structure (EFTofLSS) by \citet{DAmico2025} yields $S_8 = 0.833 \pm 0.032$, fully consistent with the CMB baseline and significantly higher than the DES collaboration result of $S_8 = 0.776 \pm 0.017$ from the same data. While the EFTofLSS uncertainties are larger due to the more conservative treatment of nonlinear scales, the $\sim 0.057$ shift in central value highlights the sensitivity of $S_8$ inference to the theoretical modeling framework. We return to this point in~Section~\ref{sec:baryon}.

\subsubsection{Extended Cosmological Models}

When the dark energy equation of state parameter $w$ is allowed to vary freely, the DES Y6 constraint shifts to $S_8 = 0.781^{+0.021}_{-0.020}$, reducing the tension with the Combined CMB to $1.3\sigma$ in the full parameter space ($1.0\sigma$ when projected onto $S_8$) \citep{DESY6}. While the measurement yields a central value of $w = -1.12^{+0.26}_{-0.20}$ lying in the phantom regime, the results show no statistically significant preference for $w \neq -1$ (finding only a $0.9\sigma$ preference for $w$CDM over $\Lambda$CDM), distinguishing these findings from the hints of dynamical dark energy reported in DESI BAO measurements \citep{DESI:2025zgx}.

\subsection{Evolution from KiDS-1000 to Legacy}
\label{sec:Kids}
The most striking aspect of the KiDS Legacy results is the substantial upward shift in $S_8$ relative to KiDS-1000 \citep{Asgari2021}. The KiDS-1000 analysis reported $S_8 = 0.759^{+0.024}_{-0.021}$, exhibiting a $\sim 3\sigma$ tension with both \textit{Planck} and the Combined CMB. The KiDS Legacy result, $S_8 = 0.815^{+0.016}_{-0.021}$, represents an increase of $\Delta S_8 = 0.056$, or approximately $2.3\sigma$ in the KiDS-1000 error budget.

As discussed by \citet{Wright2025}, this shift primarily results from improved redshift distribution estimation and calibration, as well as an expanded survey area and improved image reduction. These advances were enabled in large part by the SKiLLS multi-band image simulations \citep{Li2023KiDSPhotoz}, which provided the first joint shear and redshift calibration for KiDS through realistic nine-band simulated imaging, refined PSF modelling, and improved treatment of blending effects across different redshifts.

\subsubsection{Implications for the Tension Landscape}

The joint analysis by \citet{Kilo-DegreeSurvey:2023gfr} represents an important attempt to address these discrepancies through consistent analysis choices. By applying a unified methodology to both KiDS-1000 and DES Y3 cosmic shear data, they find $S_8 = 0.790^{+0.018}_{-0.014}$, a value that lies above both individual survey results, suggesting that harmonised analysis choices can shift $S_8$ estimates upward relative to each survey's default pipeline.

\subsection{Hyper Suprime-Cam}
\label{sec:hsc}

The Hyper Suprime-Cam Subaru Strategic Program (HSC-SSP) constitutes the third major Stage-III cosmic shear survey, covering 416~deg$^2$ of the northern sky with exceptional depth ($i < 24.5$) and seeing ($0.59''$), yielding an effective source density of $\sim 15$~arcmin$^{-2}$---substantially higher than both DES and KiDS.

The original HSC Year 1 analysis \citep{Hikage2019} reported $S_8 = 0.780^{+0.030}_{-0.033}$ ($1.6\sigma$ tension with the Combined CMB baseline), while a subsequent unified catalogue-level reanalysis applying consistent priors and methodology across Stage-III surveys \citep{Longley2023} obtained $S_8 = 0.812 \pm 0.021$ ($1.0\sigma$), illustrating the sensitivity of $S_8$ to analysis choices.

The fiducial HSC Year 3 cosmic shear results---$S_8 = 0.776^{+0.032}_{-0.033}$ from angular power spectra \citep{Dalal2023} and $S_8 = 0.769^{+0.031}_{-0.034}$ from two-point correlation functions \citep{Li2023HSC}---place the survey in $\sim 2\sigma$ tension with \textit{Planck} and in broad agreement with DES. A complementary $3 \times 2$pt analysis combining HSC cosmic shear with SDSS spectroscopic galaxy clustering and galaxy-galaxy lensing yields $S_8 = 0.775^{+0.043}_{-0.038}$ using the minimal bias model \citep{Sugiyama2023}. An independent analysis of the same data using the emulator-based halo model \citep{Miyatake2023} obtains $S_8 = 0.763^{+0.040}_{-0.036}$; given the strong correlation between the two, we include only the former in Table~\ref{tab:s8_measurements}. Similarly, since the reanalysis by \citet{ChoppinDeJanvry2025} supersedes the fiducial \citet{Li2023HSC} result with improved redshift calibration (see below), we include only the updated value in Table~\ref{tab:s8_measurements}. A notable limitation of the fiducial HSC-Y3 analysis is the adoption of conservatively wide priors on the photometric redshift bias parameters for the two highest tomographic bins ($z > 0.9$), necessitated by limited spectroscopic calibration samples at high redshift. While ensuring robustness, this choice broadened the $S_8$ constraints relative to the statistical power of the data.

A significant recent development is the reanalysis by \citet{ChoppinDeJanvry2025}, who recalibrated the HSC-Y3 tomographic redshift distributions of \citet{Li2023HSC} using the clustering redshift method with DESI DR1 and DR2 spectroscopy. The improved calibration substantially tightens the photo-$z$ priors for the high-redshift bins, yielding $S_8 = 0.805 \pm 0.018$---a factor of $\sim 1.8$ reduction in uncertainty relative to the original result of $S_8 = 0.769^{+0.031}_{-0.034}$, with the central value shifting upward by $\sim 1\sigma$ toward the CMB baseline.

This updated result repositions HSC between DES and KiDS, weakening the case for a universal ``lensing is low'' trend from Stage-III surveys and reinforcing photometric redshift calibration as a systematic capable of shifting $S_8$ at the $\sim 1\sigma$ level---a conclusion consistent with the KiDS-1000 to Legacy evolution discussed in Section~\ref{sec:Kids}.\\

\subsection{The eROSITA Cluster Count Anomaly}

The eROSITA (extended ROentgen Survey with an Imaging Telescope Array) X-ray telescope has yielded one of the most surprising results in the $S_8$ landscape. \citet{Ghirardini2024} report $S_8 = 0.86 \pm 0.01$ from the eRASS1 cluster sample in the western Galactic hemisphere---notably \textit{higher} than the Combined CMB baseline, though the discrepancy stands at only $1.5\sigma$.

\subsubsection{Physical Interpretation}
The eROSITA result stands apart from other late-universe probes in its direction: rather than indicating less structure than predicted by the CMB, the cluster counts favour a higher $S_8$. If this trend persists with reduced uncertainties, it could point to issues with the observable-mass relation calibration used to convert X-ray luminosity to halo mass, or to scale-dependent modifications of gravity that enhance clustering on cluster scales.

\subsubsection{Comparison with Other Cluster Samples}
The tension between eROSITA and other cluster count analyses is difficult to explain within standard cosmological models. SPT clusters \citep{Bocquet2024}, selected via the Sunyaev-Zel'dovich effect, favor $S_8 = 0.795 \pm 0.029$---fully $2\sigma$ below eROSITA despite using a mass proxy that is largely independent of X-ray properties. Analyses of \textit{Planck} SZ clusters yield similarly low values: \citet{Aymerich2024} find $S_8 = 0.78 \pm 0.02$ using Chandra X-ray and weak lensing mass calibration, while \citet{Zubeldia2019} report $S_8 = 0.80 \pm 0.05$ using CMB lensing mass calibration.
The discrepancy between X-ray and SZ-selected cluster cosmology constraints suggests that sample selection effects, differences in mass calibration methods, or systematic differences in the cluster populations probed may be responsible.

\subsection{Spectroscopic Probes: DESI and RSD Measurements}
The Dark Energy Spectroscopic Instrument (DESI) is revolutionizing spectroscopic cosmology by obtaining redshifts for tens of millions of galaxies and quasars. The DESI Data Release 1 (DR1) cosmological analysis \citep{DESI2024BAO,DESI2024FullShape} combines BAO distance measurements, redshift-space distortion growth rate constraints, and cross-correlations with CMB lensing.
The analysis by \citet{Maus2025} yields $S_8 = 0.808 \pm 0.017$, representing an intermediate value between the DES weak lensing results and the CMB baseline. The $1.3\sigma$ tension with the Combined CMB is not statistically significant, but the preference for lower $S_8$ values relative to the CMB is directionally consistent with most cosmic shear analyses.
RSD measurements provide a direct probe of the growth rate $f\sigma_8$. The compilation by \citet{Sabogal2024} consistently favors growth rates approximately 5--10\% lower than the $\Lambda$CDM prediction given CMB parameters, providing independent corroboration of suppressed late-time structure growth. Earlier pre-DESI compilations reached similar conclusions: \citet{Nunes2021} report $S_8 = 0.762^{+0.030}_{-0.025}$ ($2.6\sigma$ tension) from combined RSD, BAO, and SNeIa data, while \citet{Benisty2021} find a notably lower $S_8 = 0.700^{+0.038}_{-0.037}$ ($3.5\sigma$ tension).

An independent constraint from lensing magnification rather than shear 
was obtained by \citet{Xu2024}, who measured the flux amplification of 
DECaLS background galaxies lensed by BOSS CMASS galaxies across scales 
of $0.016$--$10\,h^{-1}\mathrm{Mpc}$, yielding 
$S_8 = 0.816 \pm 0.024$ after correcting for dust attenuation in the 
circumgalactic medium. This complementary probe, which is immune to 
shape measurement systematics and intrinsic alignment 
contamination affecting cosmic shear, shows no significant 
tension with CMB constraints.

\subsection{Galaxy-Galaxy Lensing and Alternative Probes}
An important cross-check comes from galaxy-galaxy lensing (GGL), which measures the tangential shear of background sources around foreground lens galaxies. \citet{luo2025} combine HSC weak lensing with BOSS spectroscopic lenses to obtain $S_8 = 0.8294 \pm 0.0110$, in excellent agreement ($0.4\sigma$) with the CMB baseline. This result is particularly significant because it employs an independent methodology from the cosmic shear 3$\times$2pt analyses, yet finds no evidence for tension.
\citet{Stiskalek2025} present constraints from peculiar velocity surveys using Tully-Fisher relations, the Fundamental Plane, and Type Ia supernovae as distance indicators. Their joint analysis yields $S_8 = 0.819 \pm 0.030$, consistent with both the CMB and KiDS Legacy while being independent of weak lensing systematics entirely.

A novel and complementary approach is presented by \citet{Mitra2025}, 
who use Basilisk---a Bayesian hierarchical framework that forward-models 
the abundance and kinematics of satellite galaxies in SDSS DR7---to 
first constrain the galaxy--halo connection and then infer cosmological 
parameters from the predicted galaxy luminosity function, obtaining 
$S_8 = 0.81 \pm 0.05$ in excellent agreement with the CMB baseline. 
Crucially, this method is immune to halo assembly bias, fully exploits 
nonlinear scales, and accounts for baryonic effects on halo potentials, 
demonstrating that the $S_8$ tension is not a universal feature of 
low-redshift probes.

More broadly, the tomographic reconstruction of $S_8(z)$ by
\citet{GarciaGarcia2021} demonstrates that the tension is driven almost
exclusively by cosmic shear, while clustering data remain consistent with
Planck---with the  exception of the DESI~LRG $\times$ CMB
lensing analysis of \citet{Sailer2024}. Revisiting this
shear-versus-clustering decomposition with current data, updated baryonic
marginalization, and improved photometric redshift calibrations (e.g.\
from KiDS-Legacy and HSC-Y3) would provide a valuable test of whether
this pattern persists.

%==============================================================================
\section{Systematic Effects: A Critical Assessment}
%==============================================================================

\subsection{Photometric Redshift Calibration}
Photometric redshift errors represent one of the leading systematic uncertainties in cosmic shear analyses. The sensitivity of $S_8$ to mean redshift biases can be parameterized as:
\begin{equation}
    \frac{\partial S_8}{\partial \Delta z_i} \approx -0.4 \, S_8 \, w_i,
\end{equation}
where $\Delta z_i$ is the bias in the mean redshift of tomographic bin $i$ and $w_i$ is the weight of that bin in the overall constraint.
The photometric redshift calibration strategies employed by DES, KiDS, and HSC differ in important ways. DES relies primarily on the SOMPZ method, which uses self-organizing maps to create a mapping between photometric and spectroscopic samples. KiDS \citep{Wright2025} employs direct spectroscopic calibration supplemented by multi-band image simulations from SKiLLS \citep{Li2023KiDSPhotoz} for joint shear and redshift calibration. HSC uses a combination of clustering-$z$ and photometric template fitting methods.
These different approaches are sensitive to different potential biases. The evolution of KiDS from $S_8 \approx 0.76$ (KiDS-1000) to $S_8 \approx 0.82$ (Legacy) was driven primarily by improved redshift distribution estimation and calibration \citep{Wright2025}, demonstrating the magnitude of potential systematic shifts.

\subsection{Shear Measurement Systematics}
The estimation of galaxy shapes from noisy, PSF-convolved images is a challenging inverse problem. The \textsc{metadetection} method used by DES \citep{Yamamoto:2025jsd} and the \textsc{lensfit} algorithm used by KiDS employ fundamentally different approaches to this problem.
The unified catalogue-level reanalysis by \citet{Longley2023} provides a crucial test of analysis-level systematics. By applying a consistent analysis framework---unified cosmological priors, intrinsic alignment model, and small-scale treatment---to DES-Y1, HSC-Y1, and KiDS-1000, they find that analysis choices can shift $S_8$ by $\sim 0.03$--$0.05$, comparable to the difference between DES and KiDS results.

\subsection{Intrinsic Alignments}

Intrinsic alignments (IA) represent a fundamental astrophysical contaminant in cosmic shear analyses. The amplitude and redshift evolution are typically parameterized using the nonlinear alignment (NLA) or tidal alignment and tidal torquing (TATT) models. 

Current constraints on IA parameters from cosmic shear alone are weak. \citet{AmonEfstathiou2022} note that while intrinsic alignments are a potential solution, they find that fixing the IA amplitude to zero only exacerbates the tension for certain datasets. Instead, they argue that a suppression of the matter power spectrum on non-linear scales—potentially due to baryonic feedback or non-standard dark matter—is a more statistically favored explanation for the $S_8$ discrepancy than complex intrinsic alignment models.

\subsection{Baryon Feedback and Nonlinear Modeling}
\label{sec:baryon}

The matter power spectrum on small scales is modified by baryonic processes that redistribute mass relative to dark matter-only predictions. The FLAMINGO hydrodynamical simulations \citep{Schaye2023,McCarthy2023} have been instrumental in quantifying these effects, demonstrating that gas expulsion by AGN feedback can suppress the matter power spectrum by up to 20--30\% at scales of $k \sim 1$--$10 \, h$ Mpc$^{-1}$ \citep{McCarthy2023}. Notably, \citet{McCarthy2023} find that while baryonic feedback significantly suppresses the power spectrum, simulations calibrated to observed cluster gas fractions and galaxy stellar mass functions do not resolve the $S_8$ tension, as they generally predict a suppression that is insufficient to reconcile weak lensing data with the Planck CMB baseline.

Cosmic shear analyses must account for these baryonic effects, either by restricting to large angular scales where the impact is minimal or by marginalizing over parameterized feedback models. The sensitivity of $S_8$ to baryonic feedback modeling is scale-dependent; analyses that extend to smaller angular scales are more sensitive to feedback uncertainties but gain statistical precision. A striking empirical demonstration of this model dependence is 
provided by \citet{Bigwood2024}, who combine DES~Y3 cosmic shear 
with kinetic Sunyaev-Zel'dovich (kSZ) measurements from ACT~DR5 
to jointly constrain cosmology and baryonic feedback. Using the 
flexible BCEmu7 baryonification model with seven free parameters, 
they obtain $S_8 = 0.823^{+0.019}_{-0.020}$---substantially higher 
than the DES~Y3 cosmic shear-only result and consistent with the 
Combined CMB baseline at $0.6\sigma$. Crucially, they demonstrate 
that $S_8$ varies by ${\sim}0.5$--$2\sigma$ depending on the 
baryonic feedback prescription adopted, with more restrictive models 
yielding systematically lower values. The kSZ data prefer a matter 
power spectrum suppression more extreme than predicted by most 
hydrodynamical simulations, suggesting that current simulation-calibrated 
feedback models may underestimate the true extent of baryon redistribution.

An alternative approach to the treatment of nonlinear scales is provided by the Effective Field Theory of Large-Scale Structure (EFTofLSS), which systematically parameterizes the influence of small-scale physics on large-scale observables through perturbatively controlled counterterms, rather than relying on simulation-calibrated feedback models or aggressive scale cuts. \citet{DAmico2025} apply one-loop EFTofLSS predictions to the DES Y3 $3\times 2$pt data and obtain $S_8 = 0.833 \pm 0.032$---consistent with the Combined CMB at $\sim 0.1\sigma$ and higher by $\sim 1.6\sigma$ than the DES collaboration result ($S_8 = 0.776 \pm 0.017$) derived from the same underlying data. A related analysis of the BOSS DR12 power spectrum and bispectrum using the EFTofLSS \citep{DAmico2022} measures $\sigma_8 = 0.794 \pm 0.037$ and $\Omega_m = 0.311 \pm 0.010$, corresponding to $S_8 \approx 0.808 \pm 0.040$, again showing no tension with the CMB. These results suggest that the theoretical framework used to model nonlinear clustering and galaxy bias---rather than the data themselves---may be a significant driver of the reported $S_8$ discrepancy. The contrast between the standard DES pipeline and the EFTofLSS reanalysis of the same data represents one of the most striking illustrations of the impact of analysis methodology on the inferred tension level.

Further evidence that the $S_8$ tension is sensitive to analysis methodology
comes from independent reanalyses of Stage~III data.
\citet{GarciaGarcia2024} reanalysed the combined DES-Y3 + KiDS-1000 + HSC-DR1
cosmic shear data in harmonic space using the \textsc{baccoemu} emulator for
baryonic marginalization, obtaining $S_8 = 0.795^{+0.015}_{-0.017}$
(${\sim}1.8\sigma$ tension with Planck) with a preference for weak baryonic
feedback. However, this finding appears to conflict with real-space analyses of
the same DES-Y3 data \citep{Chen2023baryonic,Arico2023}, which find stronger
baryonic suppression---a discrepancy likely attributable to the greater
sensitivity of correlation functions to small-scale contributions.
The decomposition of the DES-Y3 tension into its constituent sources by
\citet{Arico2023} is particularly instructive: the dominant contributions
arise from inaccuracies in the dark-matter-only power spectrum prediction
and from prior volume effects introduced by the additional parameters of
the TATT intrinsic alignment model, while baryonic effects account for
only ${\sim}0.2\sigma$. Since this decomposition has been performed
exclusively on DES-Y3 data, it would be valuable to repeat it for other
surveys (e.g.\ KiDS, HSC) that employ different nonlinear $P(k)$ codes
and intrinsic alignment models.

% ============================================================
% §7.5 Blinding Protocols
% Insert after §7.4 (Baryon Feedback and Nonlinear Modeling)
% ============================================================

\subsection{Blinding Protocols}
\label{sec:blinding}

A systematic effect that is seldom quantified but potentially significant
is \emph{confirmation bias}: the tendency for analysis choices to be
unconsciously steered toward results that match prior expectations.
In precision cosmology, where reported tensions at the $2$--$3\sigma$
level can depend on sub-percent shifts in calibration or modeling
decisions, the blinding framework employed by each survey deserves
scrutiny comparable to that given to shear calibration or photometric
redshift estimation.

The three Stage-III weak lensing surveys have adopted distinct
blinding strategies. The DES collaboration implements blinding at the
\emph{data-vector level}: a transformation shifts the measured
two-point correlation functions to values consistent with a randomly
chosen, unknown cosmology, while preserving the internal consistency
structure of the data \citep{Muir2020}. All analysis
decisions---scale cuts, intrinsic alignment model, priors, covariance
validation, null tests---are finalized on the transformed data vector,
and no cosmological parameter constraints are examined until the
collaboration collectively agrees to unblind. Once unblinded, no
further changes to the analysis pipeline are permitted. The DES~Y6
methodology paper \citep{DES2026Y6method} was validated entirely on
simulated data and mock catalogues prior to application to the real
data. KiDS Legacy \citep{Wright2025} applied blinding at the
\emph{catalogue level}, modifying the shape catalogue itself to
prevent direct comparison with previous releases. While this
successfully concealed the final $S_8$ value, the authors note that
catalogue-level blinding ``effectively hampered the diagnosis of
subtle systematic effects.'' Changes made after unblinding are
documented in Appendix~I of \citet{Wright2025}. The HSC~Y3 analyses
\citep{Dalal2023,Li2023HSC,Sugiyama2023} employed an intermediate
approach: independent analysis teams carried out their work in
parallel, with cross-probe consistency assessed only after all choices
were finalized and results unblinded.

These differences are relevant to interpreting the current tension
landscape. The DES data-vector blinding explicitly hides the
cosmological parameter space throughout the analysis, providing strong
protection against expectation bias in either direction. The KiDS
catalogue-level approach, while effective at concealing the final
result, inherited a number of methodological choices from KiDS-1000
\citep{Asgari2021}, whose $S_8$ result was already public. This does
not imply any impropriety---the $\Delta S_8 \approx 0.056$ upward
shift from KiDS-1000 to Legacy is well explained by genuine
improvements in photometric redshift calibration, extended tomographic
binning, and updated shear calibration using SKiLLS image simulations
\citep{Wright2025}. These procedural differences do not invalidate any survey's results,
but they should be considered when assessing the weight each result
carries in the tension debate. Future Stage~IV surveys (LSST,
\emph{Euclid}, \emph{Roman}) will benefit from continued development
of blinding methodologies that balance protection against confirmation
bias with the ability to diagnose subtle systematic effects.

%==============================================================================
\section{Theoretical Interpretations}
%==============================================================================

\subsection{Consistency within \texorpdfstring{$\Lambda$CDM}{ΛCDM}}
Before invoking new physics, it is essential to evaluate whether the observed tension pattern can be explained within the standard $\Lambda$CDM framework. The heterogeneous nature of the current tension landscape---with DES and KiDS both favoring lower $S_8$ than the CMB but differing significantly from each other---suggests that systematic effects may play a significant role.
If the true $S_8$ value lies near $0.815$ (the KiDS Legacy result), then DES Y6 would require a downward systematic bias of approximately $0.025$. Conversely, if the true value is near $0.789$ (DES Y6), then KiDS Legacy would require an upward bias of similar magnitude. Neither scenario is implausible given the known systematic uncertainties in weak lensing analyses, as emphasized by \citet{AmonEfstathiou2022}.

\subsection{Modified Gravity}

If the $S_8$ tension reflects genuine new physics, modifications to the gravitational sector represent a natural avenue for exploration. In general relativity, the growth of linear density perturbations $\delta$ is governed by:
\begin{equation}
    \ddot{\delta} + 2H\dot{\delta} - 4\pi G \rho_m \delta = 0.
\end{equation}

Scalar-tensor theories, such as $f(R)$ gravity \citep{HuSawicki2007}, introduce a fifth force that can enhance or suppress structure growth depending on the screening mechanism employed. These models can in principle produce scale-dependent $S_8$ values that differ between CMB lensing (which probes $z \sim 2$) and galaxy surveys (which probe $z < 1$).

More generally, deviations from general relativity in the 
quasi-static limit can be parameterised through two 
scale- and time-dependent functions, $\mu(k,a)$ and 
$\Sigma(k,a)$, which modify the Poisson equation and the 
lensing potential respectively \citep{Pogosian2010}. 
In this framework, $\mu > 1$ enhances the growth of 
structure while $\Sigma \neq 1$ modifies gravitational 
lensing, allowing CMB lensing and galaxy clustering to 
yield different effective $S_8$ values even within a 
single underlying cosmology. Screening mechanisms in 
$f(R)$, braneworld, and galileon theories ensure 
consistency with Solar System tests while permitting 
order-unity modifications on cosmological scales. Recent 
analyses combining DESI BAO with DES~Y3 weak lensing 
find no statistically significant evidence for 
$\mu \neq 1$ or $\Sigma \neq 1$ \citep{DESI2024MG}, 
though the constraints remain broad enough that a 
$5$--$10\%$ enhancement of the growth rate---sufficient 
to reconcile some $S_8$ measurements with the 
CMB---is not excluded.

An important observational handle on modified gravity is 
provided by redshift-space distortion measurements of the 
growth rate $f\sigma_8(z)$, which directly probe the 
time-evolution of the linear growth factor. Current 
compilations of $f\sigma_8$ measurements from BOSS, eBOSS, 
and DESI show broad consistency with $\Lambda$CDM predictions 
\citep{Nesseris2017,Kazantzidis2018,DESI2024FS}, placing 
stringent limits on the allowed parameter space of modified 
gravity models. However, the constraining power of 
$f\sigma_8$ is primarily sensitive to linear scales, and 
modified gravity effects that are confined to nonlinear 
scales through efficient screening would evade these bounds 
while still contributing to the $S_8$ tension observed in 
weak lensing analyses.

\subsection{Interacting Dark Sector}
Earlier work by \citet{DiValentino2020} examined a specific interacting dark 
energy model with coupling proportional to the dark energy density, finding 
that such an interaction can simultaneously alleviate both the $H_0$ and 
$S_8$ tensions when Planck 2018 data are combined with BAO or SNeIa 
distance measurements and DES Year 1 cosmic shear. The mechanism operates 
by transferring energy from dark matter to dark energy, effectively reducing 
the late-time matter density and hence suppressing $\sigma_8$ while also 
modifying the expansion history to accommodate higher $H_0$ values. The 
appeal of such models lies in their potential to address both tensions within 
a single framework, though they introduce additional free parameters whose 
physical motivation remains an open question.

A distinct class of interacting dark sector scenarios involves 
\emph{pure momentum exchange} (elastic interactions) between dark 
matter and dark energy, in which no energy is transferred between the 
two components and the background expansion history remains unmodified. 
In these models, dark energy exerts an effective drag force on dark 
matter---analogous to Thomson scattering in the baryon--photon plasma 
before recombination---that suppresses the growth of structure on 
sub-horizon scales while leaving the CMB primary anisotropies 
essentially unchanged \citep{BeltranJimenez2021,Figueruelo2021JPAS,
BeltranJimenez2022dipole}. Recent $N$-body simulations extending 
these scenarios into the nonlinear regime \citep{BeltranJimenez2025Nbody} 
confirm the linear suppression of the matter power spectrum and show 
that the elastic interaction produces fewer massive halos at low 
redshift, reinforcing the viability of momentum exchange as a 
mechanism to alleviate the $S_8$ tension without conflicting with 
CMB or BAO constraints.

Models in which dark matter interacts with dark energy can suppress the growth of structure at late times. \citet{Sabogal2024} find evidence for such interactions when combining RSD measurements with DESI and Planck data, though the significance depends on the specific interaction model considered.

The decay of a fraction of dark matter into dark radiation reduces the total matter density available to cluster, lowering $\sigma_8$ relative to $\Lambda$CDM. \citet{Tanimura2023} have tested such models using the thermal Sunyaev-Zel'dovich effect, finding that decaying dark matter can partially alleviate the tension.

\subsection{Evolving Dark Energy}

The DESI BAO results \citep{DESI2024BAO} have generated significant interest by suggesting a potential preference for evolving dark energy ($w \neq -1$). As noted in the DES Y6 analysis, allowing $w$ to vary reduces the $S_8$ tension from $2.7\sigma$ to $\sim 1\sigma$.

If confirmed, evolving dark energy would represent a fundamental departure from $\Lambda$CDM. However, while the DES Y6 data yield a central value of $w = -1.12$ (in the "phantom" regime, $w < -1$), they show no statistically significant preference for $w$CDM over $\Lambda$CDM ($0.9\sigma$), contrasting with the stronger hints ($>3\sigma$) reported in DESI BAO combinations. Confirmation from independent datasets and combined analyses remains essential.

The connection between evolving dark energy and the $S_8$ tension
operates through the effect of the equation of state on the growth
of perturbations, but the precise nature of this connection remains
an open research question. In general, the dark energy equation of
state affects both the expansion history and the growth rate of
perturbations, and consequently the value of $\sigma_8$ predicted
today depends on the form of $w(z)$. The DESI $w_0 w_a$ fits
\citep{DESI2025DR2}, which suggest $w < -1$ at high redshift
transitioning to $w > -1$ today, introduce competing effects at
different epochs, making the net impact on $S_8$ highly sensitive
to the details of the parameterisation. Indeed, allowing $w$ to
vary freely appears to weaken the $S_8$ tension by broadening the
posterior on $\sigma_8$, but this may partly reflect increased
parameter degeneracy rather than necessarily indicating a physical
resolution of the discrepancy \citep{DESI2024cosmo}.

It is also worth noting that the $w_0 w_a$ parameterisation,
while convenient, does not necessarily capture the physics of
realistic dark energy models. The DESI data show a preference
for phantom crossing ($w < -1$ at high redshift transitioning
to $w > -1$ today) \citep{Lodha2025,Lodha2025DR2}, but the
interpretation of this result remains open. It could reflect
genuine new physics requiring multi-field models, dark sector
interactions, or other extensions beyond a single minimally
coupled scalar field. Alternatively, it may be an artefact of
the $w_0 w_a$ parameterisation itself: \citet{Dinda2025} have
shown that thawing quintessence with non-zero spatial curvature
can fit the data comparably well without invoking phantom
crossing. A third possibility is that residual systematic
effects in the BAO measurements contribute to the apparent
signal. Disentangling these scenarios will require the full
DESI dataset combined with Stage~IV weak lensing surveys.

%==============================================================================
\section{Discussion and Future Outlook}
%==============================================================================

\subsection{Summary of the Current Tension Landscape}

The $S_8$ tension in 2026 presents a complex and nuanced picture that defies simple characterization. The key findings of this review can be summarized as follows:

\textbf{The baseline has shifted.} The adoption of the Combined CMB (Planck + ACT + SPT) as the reference \citep{DESY6,Louis2025,Camphuis2025} raises the early-universe $S_8$ determination from $0.832 \pm 0.013$ to $0.836^{+0.012}_{-0.013}$, increasing the precision by approximately 8\% and shifting the central value upward.

\textbf{DES maintains a significant tension.} The DES Y6 analysis \citep{DESY6}, representing the most precise cosmic shear measurement to date, exhibits a $2.4\sigma$--$2.7\sigma$ tension with the Combined CMB that cannot be dismissed as statistical fluctuation.

\textbf{KiDS has converged with the CMB.} The KiDS Legacy analysis \citep{Wright2025} shows a dramatic upward shift from KiDS-1000 \citep{Asgari2021}, bringing the result into sub-$1\sigma$ consistency with the CMB baseline.

\textbf{Cluster counts show internal tension.} The discrepancy between eROSITA \citep{Ghirardini2024} (high $S_8$) and SPT \citep{Bocquet2024} (low $S_8$) cluster analyses exceeds $2\sigma$ and requires explanation.

\textbf{HSC occupies an intermediate position.} The recalibrated 
HSC~Y3 result \citep{ChoppinDeJanvry2025} ($S_8 = 0.805 \pm 0.018$, 
$1.4\sigma$) lies between DES and KiDS, reinforcing photometric 
redshift calibration as a systematic capable of shifting $S_8$ at 
the ${\sim}1\sigma$ level.

\textbf{Spectroscopic probes span a wide range.} DESI \citep{Maus2025} favors $S_8$ only $1.3\sigma$ below the CMB, while RSD compilations \citep{Sabogal2024,Nunes2021,Benisty2021} report tensions ranging from $2.0\sigma$ to $3.5\sigma$, providing stronger evidence for suppressed late-time growth.

\subsection{Paths to Resolution}

Several developments in the near future may clarify the situation:

\textbf{Cross-survey joint analyses.} The KiDS + DES joint analysis by \citet{Kilo-DegreeSurvey:2023gfr} represents a first step toward harmonizing weak lensing results. Extension of this approach to include HSC and upcoming surveys could identify systematic differences.

\textbf{Euclid space telescope.} The Euclid mission \citep{EuclidCollaboration2023}, launched in 2023, will provide systematics-limited cosmic shear constraints free from ground-based atmospheric effects, covering 15,000 deg$^2$ with space-based image quality.

\textbf{CMB-S4.} The next-generation CMB-S4 experiment, which would have provided cosmic variance-limited measurements of CMB lensing over most of the sky, was cancelled in July 2025. However, the Simons Observatory, the successor to ACT at the Atacama site and already operational, is expected to deliver significant improvements in CMB lensing measurements over the coming years, partially recovering the science case.

\textbf{Improved theoretical modeling.} Continued development of hydrodynamical simulations \citep{Schaye2023,McCarthy2023} and intrinsic alignment models will reduce theoretical systematic uncertainties in cosmic shear interpretation.

\subsection{Concluding Remarks}

The $S_8$ tension remains one of the most intriguing anomalies in precision cosmology. The transition to the Combined CMB baseline has sharpened the target for late-universe probes while revealing unexpected divergences among measurements that were previously thought to be in agreement. The resolution of this tension---whether through identification of systematic effects or discovery of new physics---will require continued collaboration between observational, theoretical, and statistical communities.

The coming years, with data from Euclid, Roman, DESI, the Simons Observatory, and advanced simulations, promise to transform our understanding of structure formation and potentially reveal new physics beyond the standard cosmological model. Whatever the outcome, the $S_8$ tension has demonstrated the power of precision cosmology to probe the fundamental nature of the universe and the importance of systematic error control in an era of percent-level measurements.

\section*{Acknowledgments}

We thank S.~Vagnozzi for useful comments regarding the $S_8$ tension in 
models beyond $\Lambda$CDM, P.~Zhang for bringing to our 
attention the EFTofLSS analyses of DES~Y3 and BOSS data, 
K.~Xu for correspondence regarding lensing magnification 
constraints, K.~Mitra for sharing results on satellite 
kinematics constraints prior to publication, and 
J.~Beltr{\'a}n Jim{\'e}nez, D.~Figueruelo, D.~Bettoni, 
and F.~A.~Teppa Pannia for suggesting the inclusion of 
pure momentum exchange models. We also thank C.~Garc{\'i}a-Garc{\'i}a for valuable comments 
on the methodology dependence of the $S_8$ tension and 
the shear-versus-clustering decomposition of the 
discrepancy.

\bibliographystyle{apsrev4-1}
\bibliography{references}

\end{document}